# Voice Search and Typed Search Performance Comparison on Baidu Search System

Peking University

Hanqing Huang 1500018610

Kezia Irene 1802010110

Nahyun Ryu 1802010374

## Abstract

While the voice search system is getting more and more developed, some people still have difficulties in searching for information with voice search technology. This paper is a pilot study to compare the search performance of people using voice search and typed search using Baidu search system. We surveyed and interviewed 40 Chinese students who have been using the Baidu search system. Afterward, we analyzed 8 people who had a middle to advanced searching ability by their behaviors, search results, and average query length. We found that there are a lot of variations among the participants' time when searching for different queries, and there were some interesting behaviors that were a number of participants. We conclude that more participants are needed to make a firm conclusion on the performance comparison between voice search and typed search.

*Keywords:* voice search, typed search, search performance

## Introduction

In modern society, we cannot separate ourselves from information retrieval. As people look for convenience more and more, many technological developments have been accomplished and voice search technology is also one of them. 20% of mobile queries of Google search system are voice searches. Moreover, an article titled "Voice Search Trends: You Need To Stop Talking And Start Listening" said that half of the searches will be performed with voice, and for the last few years, 40% of the searches done by adults were voice-based. Also, according to "Voice search and chatbots are transforming commerce" shows the main reasons people using voice search with their mobiles. 43% of participants said that it is quicker than going on a website or using an app, and 42% said that they can search something while driving or at other times when they cannot interact with their handset. With this technology, we can ask the search system to play music, navigate to a destination, forecast weather or do more simple tasks. Now, users do not have to type a word in search box directly for searching information anymore. Instead, just by clicking the microphone icon and say what they want to know is enough for searching. It can reduce users' efforts and time to type the words. It is absolute that voice search technology is the one that allows users to find information more conveniently.

However, despite these advantages, in real life, some people still use text search method more often. People mainly use voice search for simple tasks which do not require queries more than 2. It evoked curiosity about how voice search system actually performs for complex search tasks compared to text search.

## Reviews on Related Studies

1. Jansen B. J (2007) Categorized users intend into an informational, navigational, and transactional search. The result is that they were able to define which queries belong to each category, and they found out that from the random sampling, 80.6% queries were informational queries, 10.2% queries were navigational queries, and 9.2% queries are transactional queries. In conclusion, they give suggestions to search engine, that if they want to improve, they have to know their user's behavior, especially efforts to understand the underlying intent of the searchers. The higher percentage of information queries indicates that users view search engines primarily as information retrieval tools rather than instruments of navigation or commerce.

2. Jiang J., Jeng, Wei (2013) study the kind of error that often happened on English voice search. This study suggested that voice input errors as the essential issue to be resolved in voice search. A possible solution is to better support users' query reformulation, which includes designing better interface supporting voice query reformulation and developing query suggestion algorithms using both lexical and phonetic information.

3. Sa, N (2016) study found that the main problem with voice search is user can't reformulate their queries easily. In this research, it is hypothesized that the user's search experience will be improved in the experimental system with the new modification feature than in the baseline system and the user's search behavior in the experimental system will be different from that of the baseline system.

4. Chau, M., Fang, X., & Yang, C. C. (2007) found that Chinese queries will be harder to be analyzed by the search engine compared to English. The researchers found out that the co-occurrences of terms in Chinese queries are pretty different from English queries. For example the character usage, in Chinese, there is no space between terms, so it is not easy to tokenize two terms automatically without advanced algorithm.

5. According to Ying and Sholer(2007), rather than simply suggesting a correlation with AP, analyzing the users' search performance is much more essential to show the actual difficulty that users face when searching on different topics.

6. Sarhan(2014) proved that voice search engine needs more training and testing to make it more accurate on catching users' queries and searching adequate information that can satisfy users' information needs.

7. According to the paper by Ong, Järvelin, Sanderson, and Scholer (2018), the searchers, data, and logging should be included to the methodological part, which can be helpful for our project to design the research method based on the good points and limitations of the study.

8. Turpin and Scholer (2006) categorized the factors that might affect the result, which could be a reference source for our data analyzing step. By picking out some factors that can influence results before starting analysis of data in eartnest, we might be able to find out the causes and effects more analytically.

9. Liu, Li and Yang(2017) made a review about the research progress in interactive information retrieval abroad. This article can reveal the specifics of language search we need to pay attention to because Chinese articles on interactive information retrieval are still relatively small.

10. Azzopardi and Brennan(2013)'s article revealed how query cost affect search behavior. In our experiment, the voice search and the type search may have different query costs. The NASA TLX questionnaire they gave to the participants inspired our interview design.

11. In all type of search, the user may feel a delay that can also be attributed to the delay of the query time in this article, which will impact the user's behavior on research(Maxwell & Azzopardi, 2014). When we compared voice search and type search, some users believed that voice search is faster than type search. So there is a "delay" in type search.

12. Guy(2016) find that the difference in time between voice and text queries in a day is significant during the day, voice search is more used in daytime, and type search are more used at night. He finds that the average length of a session in voice search and text search are similar. Using one word as query is more rare in voice search that text search.

## Methods

### Participants

Eight Chinese students from different majors that have novice text search ability and beginner voice search ability.

| Number | Gender | Department | Grade |
|--------|--------|------------|-------|
| 1. | Male | Guanghua School of Management | 2018 |

| 2. | Female | School of Foreign Languages | 2018 |
| 3. | Male | National School of Development | 2016 |
| 4. | Female | Journalism and Communication | 2016 |
| 5. | Male | School of engineering | 2015 |
| 6. | Female | School of Languages | 2017 |
| 7. | Female | School of Urban Planning | 2017 |
| 8. | Male | School of Information Management | 2017 |

**Assessments and Measures**

1. Total websites visited: in the given 10 minutes time, the number of websites opened by participants is counted
2. Average time spent per query: count the average time that is spent on one query (include opening websites and scrolling through query)
3. Query reformulation: count how many times participants reformulate their queries (only changing one word from the query)
4. Repeated Query: count how many queries being searched more than once
5. Average Query length: count the average query length that is searched by the participants
6. Strange behavior: Track the strange participants' behavior. Such as opening video results, wrong voice recognition, forgot the rules by accident and adding or subtracting the word one by one

**Experimental Procedure**

1. Ask 30 people to answer a questionnaire about their searching ability (1-5)
2. Among them choose who rate them as between 3-5 (about 8-10)
3. Ask participants to sign consent paper about the experiment
4. Ask participants to find information about queries with 4 tasks using Baidu mobile search engine. 2 of the tasks are in an illness context, where participant needs to find medicine to cure their friends. And the other 2 are in a food-making context, where participant needs to make a recipe for a meal. And we make one of the illness tasks into a voice search task, and the other one into a type search task. Same for the food tasks. The details of the task are in Appendix A.

5. Interview the participants. Ask the difference they felt between type search and voice search and what can be improved in voice search. The details of the interview questions are shown in Appendix B.

6. Analyze the data collected. Compare the length of the average queries between voice search and type search. And we will mark every query by "natural language" or "not a natural language", and count them, compare them between voice search and type search.

**Data Collection Instruments:**

1. An android phone and screen recorder application
2. Voice recorder for an interview after the experiment

<p style="text-align:center">**Results and Discussions**</p>

We do the analysis based on the form we count. We first count the number of queries, the total search time, total documents visited, amount of repeated queries they used, amount of queries they reformulate. Then we calculate the average time spent per query.

**Outcome 1 Qualitative Data Analysis**

*Average amount of queries per task*. Unfortunately, there's no significant difference between the average amount of queries they used for each task. The average queries for voice search are around 8.625, and the average queries for text search are around 8.375.

*Average time spent per query.* Also, there's no significant difference between voice and text search on average time spent per query. In the illness task, people use more time when they are doing type search, but in the food task, people use more time when they are doing voice search. The inconsistent results on average time spent per queries with a huge standard deviation indicate there is a factor unknown driven by the difference of each participant.

   ***Query reformulation and query repetition.*** Both voice search and text search do a similar amount of query reformulation and query repetition.

|  | illness-voice | food-voice | illness-type | food-type |
|---|---|---|---|---|
| Average queries | 8.875 | 8.375 | 7.125 | 9.625 |
| SD | 2.90 | 4.138 | 2.47 | 1.768 |
| Average time spent/ query | 57.7 | 66.5 | 76.7 | 52.1 |
| SD | 24.51 | 28.93 | 33.10 | 12.58 |
| Average Query Length | 7.4 | 5.4 | 7.915 | 4.6 |
| SD | 1.97 | 0.99 | 1.94 | 2.11 |

   ***Average query length.*** It is clearly shown that the average query length of the illness task is longer than the menu task. But there is no significant difference on average query length between voice search task and text search task (means<1).

   We do not find any significant difference between voice search and text search in terms of average query length, total query per time given, average time spent per query because of data inconsistency with a big standard deviation between participants. Query reformulation and query repetition do not play a significant role either. A further research that considers different aspects that could make a difference between voice search and text search is needed.

**Outcome 2 Natural Language or Multiple Terms**

   The queries were classified into a *natural language* style or a *multiple terms* style. We made a hypothesis that people may prefer using *natural language* style in voice search task, and *multiple terms* style in the type search task. *Natural language* style refers to people forming queries using a whole sentence, like people asking questions in the real world. *Multiple terms* style refers to people forming queries using multiple keywords.

Among all 8 participants, there's only one participant who prefers to use a natural *language* style in both type search task and voice search task. She reported in her daily life, she performs all queries in *natural language* style too. The other seven participants prefer to use a *multiple terms* style. In the interview, they report that because they are used to use *multiple terms* style in typed search in daily life, so even in the voice search, they form the queries using keywords.

The results reveal that the style people use to form queries may mostly depend on what they are used to in daily life. Since all of our participants rate their search skills above average, it is reasonable that they are not switching their style in all the search tasks.

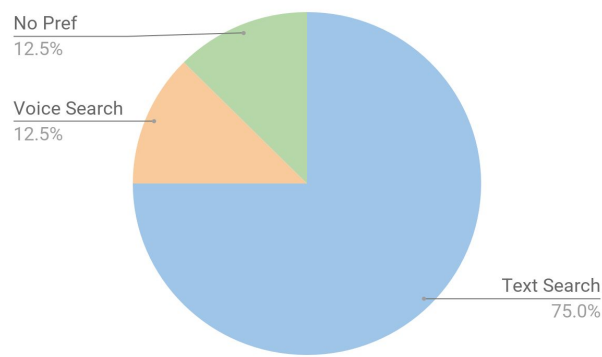

There's also one interesting result. Participant 6, who used natural language in both type search and voice search, is the only one who prefers voice search over text search. There are two possible reasons. Either voice search is a better fit for natural language style query inherently, or voice search is optimized for natural language query.

But the sample size is too small (only 1 participant used natural language search in voice search), to make any conclusion based on this finding. The hypothesis can be tested in the future works.

**Outcome 3 The behaviors**

When analyzing the results, some interesting behaviors caught our attention.

***Visit the same websites / queries multiple times.*** It might be participants want to check if there were better results in other documents/queries, but found the former one is the better, so they revisited the former one. In the mobile search app we used for conducting experiments, Only one website can be opened at the same time, so documents can't be saved for later examination. That might be another reason for using the same queries again and again.

***Adding / subtracting words one by one.*** This behavior is mostly observed in illness tasks, participants add symptoms one by one, or put all symptoms in the query and start to subtract them. That might happen when the participants didn't get the results they expected.

***Repeated forming queries in the voice search task.*** Often, the system cannot get some words correctly, so participants have to repeat their queries many times. For example, when some of the participants want to search "突尼斯", the system consistently recognized the input voice as "toons", so the participants need to repeat "突尼斯" in order to get a correct query.

## Outcome 4 The advantage and disadvantage of the Voice Search

We did an interview after the participants complete all 4 tasks. The participants reported their feeling of the voice search which they experienced in the experiment.

The advantages are as follows: The voice search is relatively easy and fast. Because it seems using it saves the time of typing.

There are four disadvantages. First, users cannot input space and other punctuations in voice search, the system automatically recognizes the voice input as a whole sentence. Participants complained that it affected their search results. Second, the system failed to recognize some words, such as "突尼斯" as we mentioned before. Part of the reason is that some participants have an accent which confused the system. But a good system may accommodate for more accents.  Third, when conducting the voice search, it is time-consuming to add other words base on the previous queries. In order to do so, users need to repeat the whole query again before adding new words. But it is easily achieved in typed search by just typing one more word. Finally, one participant reported that when she was conducting voice search, it is hard for her to think as thoroughly as performing type search. Users have to form queries quickly in voice search, otherwise, they have to input all over again. Thus, it may be concluded that when doing a complex task which involves intensive thinking, type search may be a better choice.

## Limitation and Future Work

With recent voice commanding technology development, it is meaningful to conduct the study that aims to compare voice search and text search, which is already served as one of traditional technology. However, there are still some limitations in the study, which should be resolved in the future related research.

Firstly, we missed some possible mediation factors. We could not control the time that participants write information they have searched for required tasks. It might affect the research result. Moreover, Since two out of three of our experimenters are not good at Chinese, we do the query length count by word instead of character, that may cause different experiment results. Lastly, our participants are all undergraduate students from Peking University, and they are all good at search skills. The good search skills might be affect their preference to search method. They are too used to the traditional type search. If the participants are older people or school kids, the experiment results may change.

Second, there were unexpected search behaviors. We designed the illness task with random symptoms, so there is not a correct answer and participants were expected to form more queries in 10 minutes. But in the real circumstance, some of the participants spend too much time on checking the documents to get the "right" answer, instead of making queries. Also, we cannot mark the outcome of each task, since there are no correct answers. Thus it is hard to compare the completeness of the each search. Plus, outliers data should be excluded from the calculation.

For the future study, it might be able to control mediation factors by dividing each task into a search part and a write-down answer part. Also we can ask people to use screenshots or other technical devices to record the answer, so that they can review and finish their answers after the whole searching task. Moreover, we should include more various groups of people and make more general and realistic conclusion. For designing search tasks more adequate to our research question, the query tasks should be developed with certain correct answer. One of the participants suggested that the voice search is more used in the Q&A context, and expected to get a answer in voice too. Since our experiment is more focused on the complex task, it can be a future work in testing how the voice search with the chatbot performed on the complex task.

**Appendix A   Tasks**

1. Voice search tasks:

    a. Kim lives in a remote village in the Amazon rainforest area. She is a worker and was poorly educated. One day Kim gets really sick, and unfortunately the local medicines there cannot cure her. Her nose bleeds often and she coughs a lot.  She feels pain on her stomach sometimes. Kim is not able to go to the hospital because of the location is too far and her village doesn't have any transportation to go with. But luckily, a nearby village has a drug store that can help heal Kim's disease. However, Kim still doesn't know what kind of disease that she has and what medicine she should buy. As Kim's friend who knows how to use a search engine, you want to help Kim by searching what disease that she has and what medicine she should buy.

    b. You want to invite an Indian friend, a British friend, a Japanese friend, and a Tunisian friend to come to the house for a Christmas Party. You want to be able to make a meal that makes them happy, and they are all really missing home. You also want them to taste traditional Chinese food. You are going to buy the materials in the shop. Please prepare your dishes and the recipes, so that you get a shopping list.

2. Text search tasks:

a. Ben lives in a remote village in Papua New Guinea. He works as a farmer and was poorly educated. One day Ben gets really sick, and unfortunately the local medicines there cannot cure him. His head feels so dizzy and his vision blurred. He feels pain on his back and his tummy is bloated. Sometimes Ben even coughing out blood. Ben is not able to go to the hospital because of the location is too far away and his village doesn't have any transportation to go with. But luckily, a nearby village has a drug store that can help heal Ben's disease. However, Ben still doesn't know what kind of disease that he has and what medicine he should buy. As Ben's friend who knows how to use a search engine, you want to help Ben by searching what disease that he has and what medicine he should buy.

b. You want to invite an Indonesian friend, an American friend, a Korean friend, and an Egyptian friend to come to the house. You want to be able to make a meal that makes them happy. At the same time, your American friend is still losing weight, so he hopes that this meal will not be high in calories. Also, Your Egyptian friend is a Muslim. Your Korean friend really hates carrots. Please prepare your dishes and recipes.

**Appendix B   Interview Questions**

1. Which one do you prefer, the voice search or the type search? And why?
2. What do you think is the difference between voice search and type search?
3. What do you think can be done to make BAIDU voice search engine better
4. Is there any suggestion for our task designs?